\documentclass[a4paper,11pt]{article}
\usepackage{jheppub} % for details on the use of the package, please
\usepackage{graphicx}
\usepackage[figurename=Fig.]{caption}
\usepackage[tablename=Tab.]{caption}
\usepackage{color}
\usepackage{float}
\usepackage{amssymb}
\usepackage{amsmath}
\usepackage{mathrsfs}
\usepackage{textcomp}
\usepackage{bbm}
\usepackage{cases}
\usepackage{makecell}
\usepackage{pdflscape}
\usepackage{subfigure}
\usepackage{pdfpages}
\usepackage{makeidx}
\usepackage{bm}
\usepackage{xstring}
\usepackage{lipsum}
\usepackage{slashed}
\usepackage{multicol}
\usepackage{setspace}
\usepackage{booktabs}
\usepackage[table]{xcolor}
\usepackage{multirow}
\usepackage{braket}
\usepackage[separate-uncertainty]{siunitx}
\usepackage[displaymath, mathlines]{lineno}

\newcommand{\sims}{\raisebox{0.5ex}{\texttildelow}}

\newcommand{\vo}{\vec{o}\@ifnextchar{^}{\,}{}}
\def\up{\mathrm}

\renewcommand{\thetable}{\Roman{table}}

\graphicspath{{PDF2/}}

\def\slash#1{\setbox0=\hbox{$#1$}           % set a box for #1 
   \dimen0=\wd0                                 % and get its size 
   \setbox1=\hbox{/} \dimen1=\wd1               % get size of / 
   \ifdim\dimen0>\dimen1                        % #1 is bigger 
      \rlap{\hbox to \dimen0{\hfil/\hfil}}      % so center / in box 
      #1                                        % and print #1 
   \else                                        % / is bigger 
      \rlap{\hbox to \dimen1{\hfil$#1$\hfil}}   % so center #1 
      /                                         % and print / 
   \fi}                                         %     
\def\sl#1{\setbox0=\hbox{#1} 
  \dimen0=\wd0 
  \rlap{\hbox to \dimen0{\hss/\hss}}% 
  #1}

\title{Trend and forecasting of the COVID-19 outbreak in China}

\author[1]{\large{Qiang Li}\note{liruo@nwpu.edu.cn}}
\author[2]{\large{Wei Feng}\note{wfeng@xidian.edu.cn}}

\affiliation[1]{School of Physical science and Technology, Northwestern Polytechnical University, Xi’an, 710129, P. R. China.}
\affiliation[2]{School of Electronic Engineering, Xidian University, Xi’an, 710071, P. R. China.}

\abstract{ 
By using the public data from Jan.\,20 to Feb.\,11, 2020, we perform data-driven analysis and forecasting on the COVID-19 epidemic in mainland China, especially Hubei province. Our results show that the turning points of the daily infections are predicted to be Feb.\,6 and Feb.\,1, 2020, for Hubei and China other than Hubei, respectively. The epidemic in China is predicted to end up after Mar.\,10, 2020, and the number of the total infections are predicted to be 51600. The data trends reveal that quick and active strategies taken by China to reduce human exposure have already had a good impact on the control of the epidemic.
}

\begin{document} 
\maketitle
%\flushbottom
%\kaishu

\section{Introduction}

China, especially the Hubei province, is fighting the pneumonia epidemic by implementing various prevention and control measures\,\cite{ZhuN2019}.
Pneumonia, caused by the novel coronavirus (2019-nCoV), which may originate from the bat\,\cite{LiX2020}, is just named as the COVID-19 by the World Health Organization (WHO). The COVID-19 outbroke from Wuhan, the capital of Hubei province in China in Dec. 2019, and has spread to other provinces of China and even other countries\,\cite{Rothe2019}.
Strong human-to-human transmission is established\,\cite{CDC2020}, and there exists the risk of transportation of COVID-19 from Wuhan to 369 other cities in China\,\cite{DuZ2020}.
Until Feb.\,11, 2020, there have been 44653 cases of COVID-19 infections confirmed in mainland China, including 1113 deaths.

To prevent and control the spread of the epidemic, many strategies are needed\,\cite{WangFS2020}, and China has already taken quick and effective strategies to reduce population mobility and interpersonal contact rates, and also increase quarantine on migrants. Predicting the trend of the epidemic are quite important to the allocation of medical resources, the arrangement of production activities, and even the domestic economic development all over China.

In recent decades, two other new coronaviruses,“Severe Acute Respiratory Syndrome coronavirus  (SARS-CoV)" and ``Middle East respiratory syndrome coronavirus (MERS-CoV)", have been considered to be major epidemics worldwide. However, concerning these viruses, COVID-19 presents a higher degree of uncertainty in the scale and geographical scope of the outbreak within and outside China. 
SARS-CoV and MERS-CoV-based propagation analysis and prediction models may be no longer suitable in the fight against the COVID-19 pneumonia.
Therefore, it is very urgent to use the latest data to establish an efficient and highly suitable epidemic analysis and prediction model according to the actual situation, and then to give reliable epidemic predictions. This work could provide an important reference for the government to formulate emergency macroeconomic decisions and medical resources allocation. Moreover, the significance is also of great reference value for the country's deployment and adjustment of economic activities throughout the year 2020.

Very recently, the susceptible-exposed-infectious-recovered (SEIR) model is used to forecast the potential domestic and
international spread of this COVID-19 outbreak\,\cite{WuJT2020} with parameters estimated from other sources. The effective daily reproduction ratio and  epidemics peak are also predicted in a Ref.\,\cite{TangB2020}.
The real situation could be much more complicated and changing all the time.
Especially, with the implementation of the Chinese government's multiple epidemic control policies, the control of nationwide epidemic has become obvious. However, the medical supplies in Hubei will still affect the implementation of national policies. 
%As a result, the number of infections-time curves in different cities in China has shown extreme instability. Under such circumstances, it is urgent to establish a targeted and highly reliable epidemic situation analysis method and prediction model based on the characteristics of epidemic development in different regions, combined with government policy and medical condition analysis.
In this study, the models driven by data are used to describe the current data of the epidemic, to predict the ongoing trend, and to estimate the outbreak size of the COVID-19 in both Hubei and other areas in mainland China. This data-driven study could present another prediction to the epidemic compared with the susceptible-infectious-recovered (SIR) or SEIR model.

We will estimate the daily number and the total number of infections and deaths until the end of this epidemic, and also the corresponding turning points. This work is organized as below. In section II, we specify the data used and also shows the varies of epidemic over time. In section III, the models used to describe the data are introduced. Then in section IV, we present the final results of the predictions. Finally in section V, we give a brief summary and discussion.

\section{Data sources}
\begin{figure}[h!]
\centering
\subfigure[]   {\includegraphics[width=0.48\textwidth]{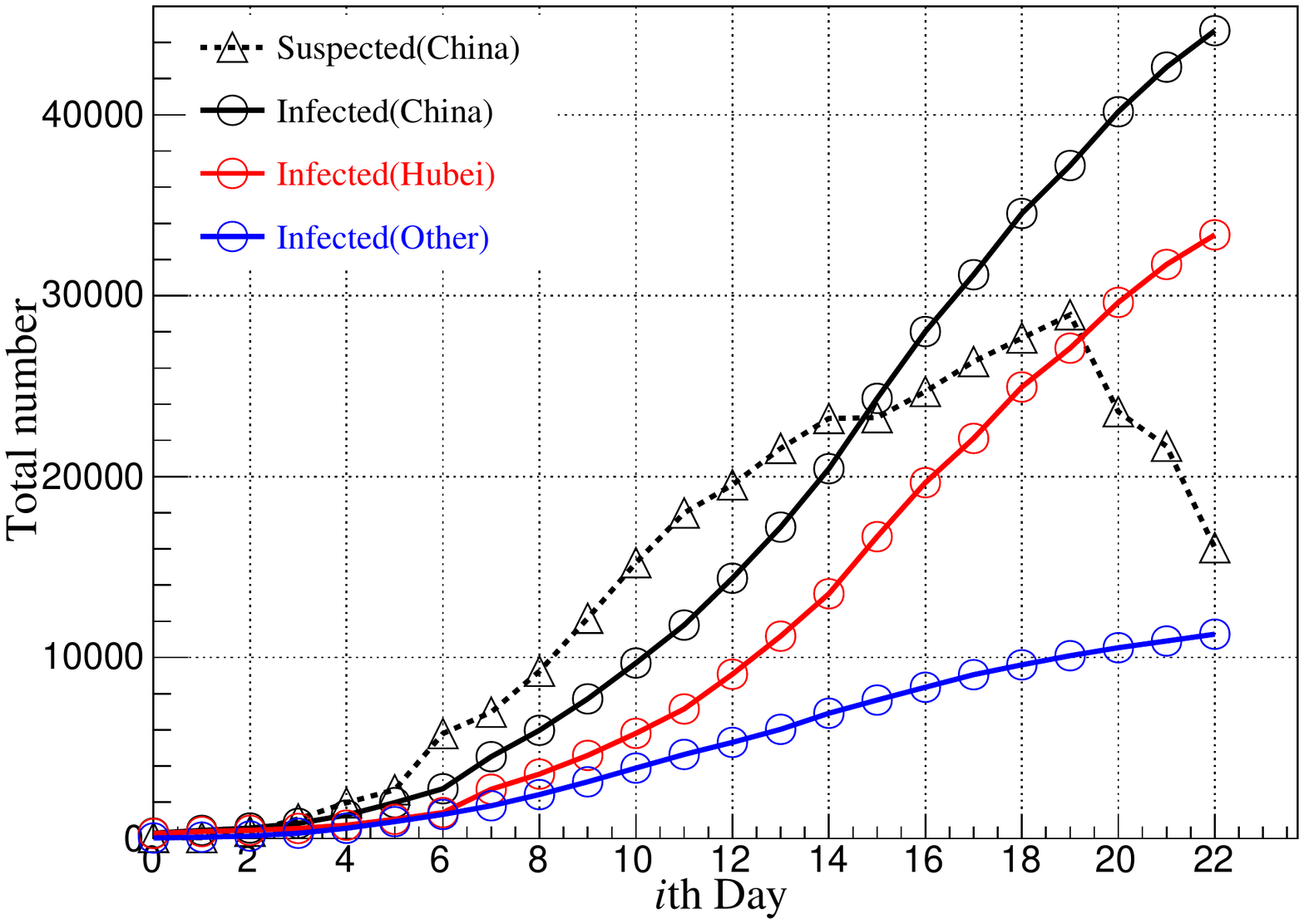} \label{Fig-I1-1}}
\subfigure[]   {\includegraphics[width=0.48\textwidth]{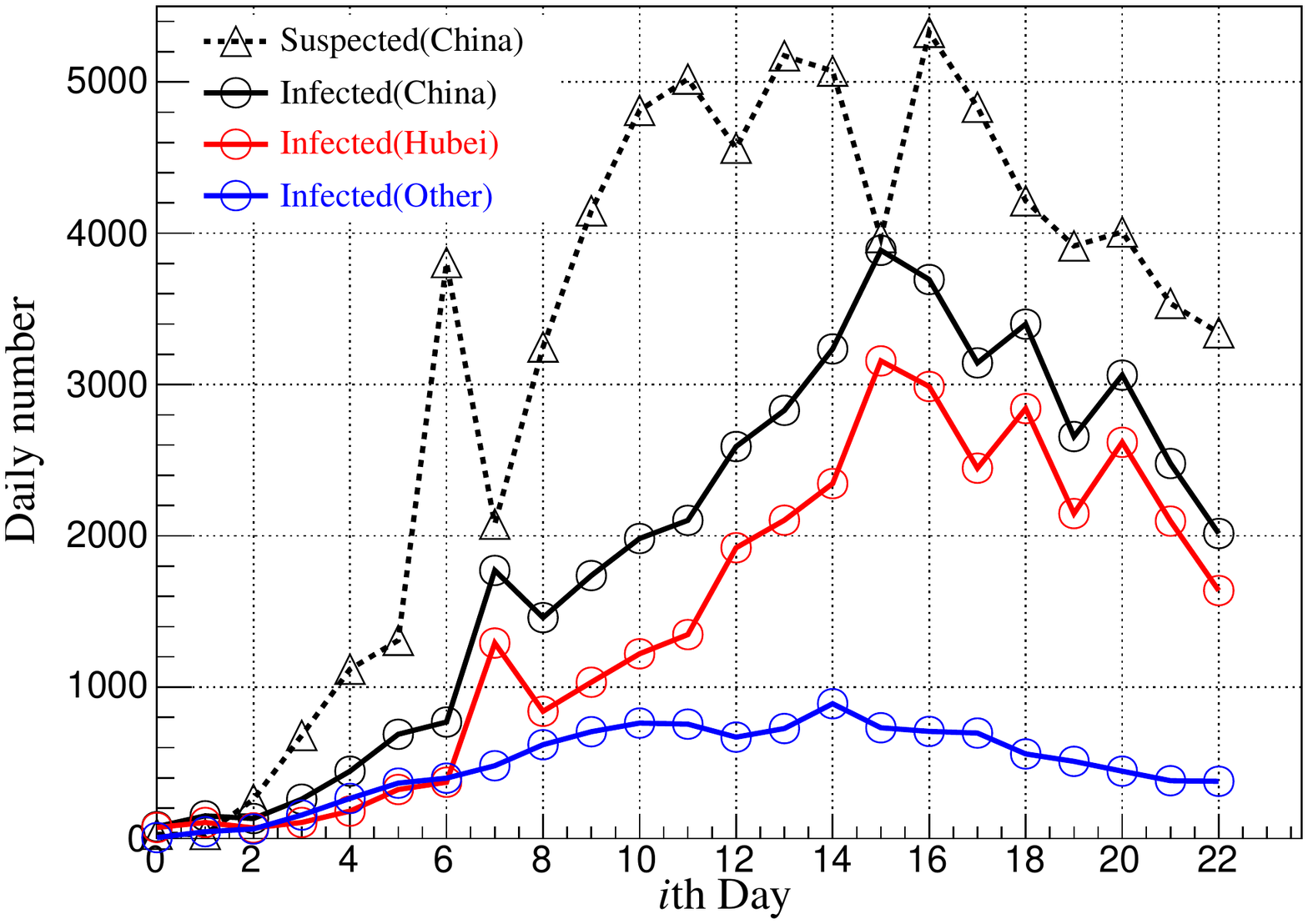} \label{Fig-I1-2}}
\caption{The number of (a) total and (b) daily suspected and confirmed cases in mainland China, Hubei, and China other than Hubei. } \label{Fig-I1}
\end{figure}

\begin{figure}[h!]
\centering
\subfigure[]   {\includegraphics[width=0.48\textwidth]{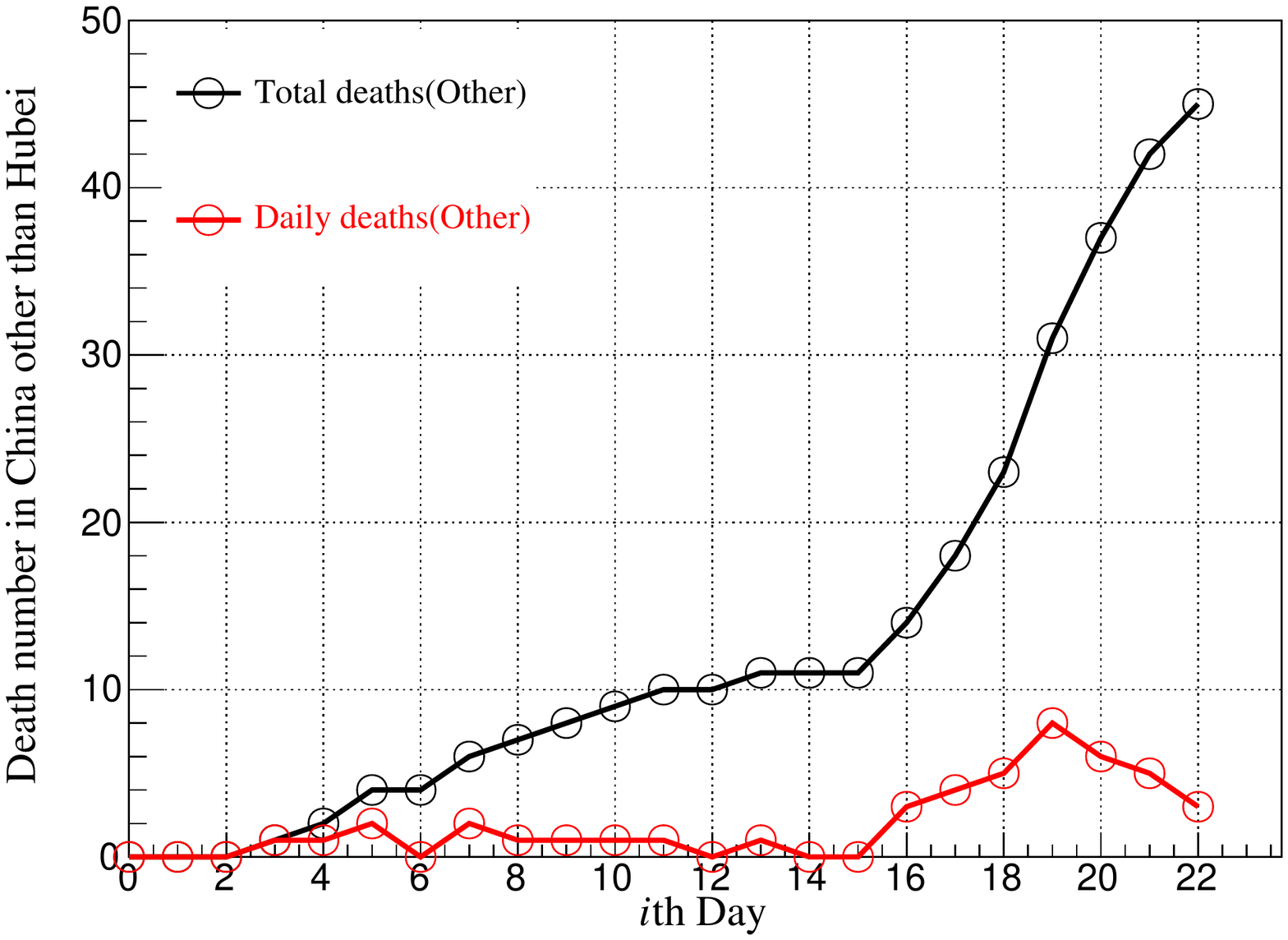} \label{Fig-I3-2}}
\subfigure[]   {\includegraphics[width=0.48\textwidth]{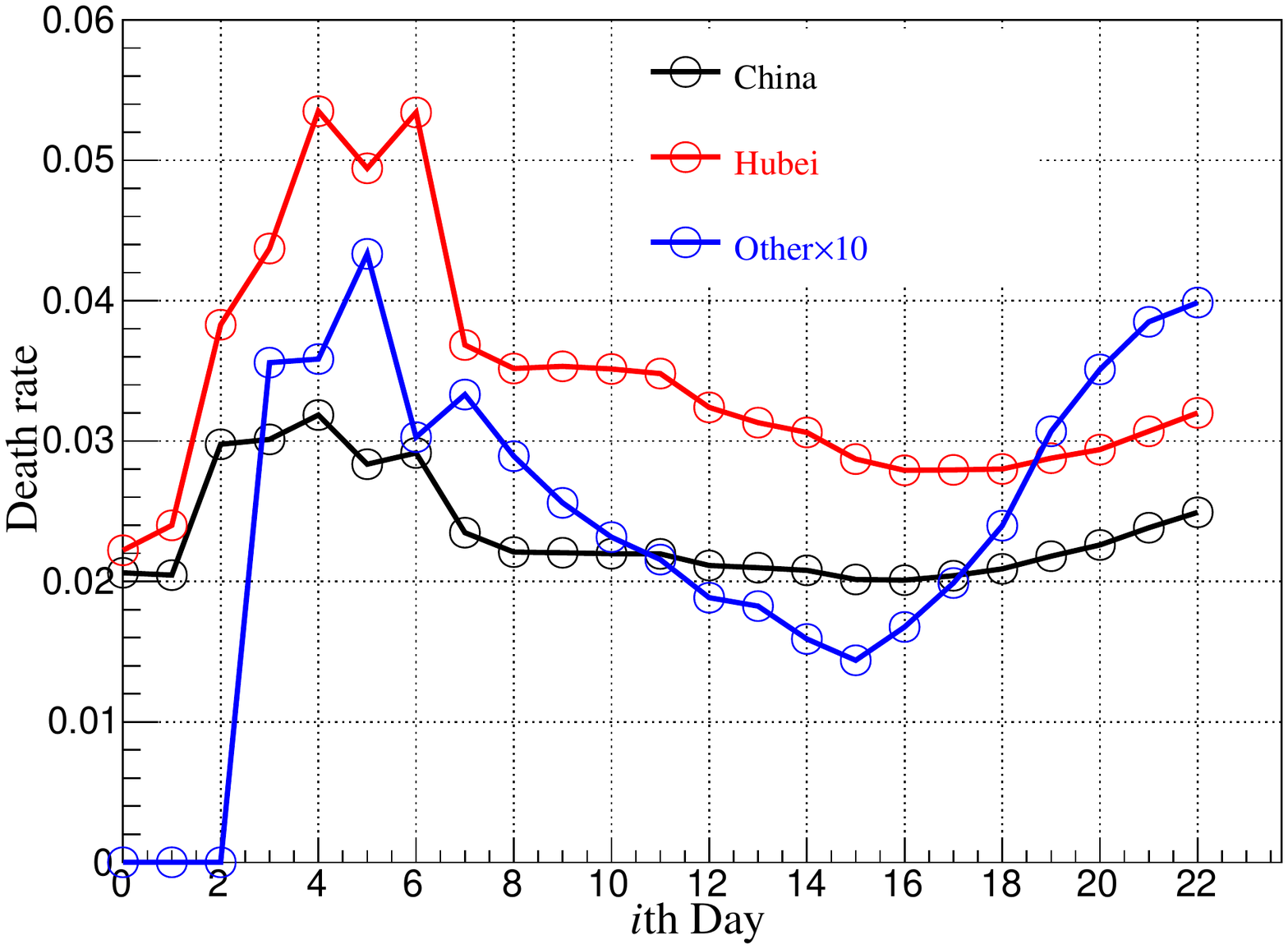} \label{Fig-I3-1}}
\caption{(b) The number of total and daily deaths over time in China other than Hubei; (a) the death rate varies over time for mainland China, Hubei, and China other than Hubei; } \label{Fig-I3}
\end{figure}

In this data-driven study, the main data used are the numbers of daily confirmed infections, totally confirmed infections, daily deaths, and total deaths.
The data used are from Jan.\,20, 2020, to Feb.\,11, 2020, where both data of Jan.\,20 and Feb.\,11 are included, reported by the National Health Commission of the Republic of China (NHC)\cite{WJW}, and Health Commission of Hubei Province (HCH)\cite{HWJW}. Jan.\,20, 2020, containing all the cases reported from 0 to 24, is the zeroth day in our research, and then others are implied. Notice the epidemic in Hubei is quite different from the other areas of China, and we will deal with the relevant data separately. All the original data used are listed in \autoref{Tab-data}, which are also graphically shown in \autoref{Fig-I1}$\sim$\,\autoref{Fig-I3}, in which “China” is used to denote the mainland China, and ``Other'' mainland China other than Hubei province. 

\begin{table}[htbp] 
  \centering
   \caption{The data of epidemic caused by the COVID-19 pneumonia in the mainland China and Hubei, including (A) the daily infections, (B) daily deaths, (C) total infections, (D) total deaths, (E) daily and (F) total suspected cases.} \label{Tab-data}
    \begin{tabular}{c|cccccc|cccc}
          & \multicolumn{6}{c}{China}                     & \multicolumn{4}{|c}{Hubei} \\
\hline
   Date       & \multicolumn{1}{c}{A} & \multicolumn{1}{c}{B} & \multicolumn{1}{c}{C} & \multicolumn{1}{c}{D} & \multicolumn{1}{c}{E} & \multicolumn{1}{c}{F} & \multicolumn{1}{|c}{A} & \multicolumn{1}{c}{B} & \multicolumn{1}{c}{C} & \multicolumn{1}{c}{D} \\
\hline
    2020/1/20 & 77    & 2     & 291   & 6     & 27    & 54    & 72    & 2     & 270   & 6 \\
    2020/1/21 & 149   & 3     & 440   & 9     & 26    & 37    & 105   & 3     & 375   & 9 \\
    2020/1/22 & 131   & 8     & 571   & 17    & 257   & 393   & 69    & 8     & 444   & 17 \\
    2020/1/23 & 259   & 8     & 830   & 25    & 680   & 1072  & 105   & 7     & 549   & 24 \\
    2020/1/24 & 444   & 16    & 1287  & 41    & 1118  & 1965  & 180   & 15    & 729   & 39 \\
    2020/1/25 & 688   & 15    & 1975  & 56    & 1309  & 2684  & 323   & 13    & 1052  & 52 \\
    2020/1/26 & 769   & 24    & 2744  & 80    & 3806  & 5794  & 371   & 24    & 1423  & 76 \\
    2020/1/27 & 1771  & 26    & 4515  & 106   & 2077  & 6973  & 1291  & 24    & 2714  & 100 \\
    2020/1/28 & 1459  & 26    & 5974  & 132   & 3248  & 9239  & 840   & 25    & 3554  & 125 \\
    2020/1/29 & 1737  & 38    & 7711  & 170   & 4148  & 12167 & 1032  & 37    & 4586  & 162 \\
    2020/1/30 & 1982  & 43    & 9692  & 213   & 4812  & 15238 & 1220  & 42    & 5806  & 204 \\
    2020/1/31 & 2102  & 46    & 11791 & 259   & 5019  & 17988 & 1347  & 45    & 7153  & 249 \\
    2020/2/1 & 2590  & 45    & 14380 & 304   & 4562  & 19544 & 1921  & 45    & 9074  & 294 \\
    2020/2/2 & 2829  & 57    & 17205 & 361   & 5173  & 21558 & 2103  & 56    & 11177 & 350 \\
    2020/2/3 & 3235  & 64    & 20438 & 425   & 5072  & 23214 & 2345  & 64    & 13522 & 414 \\
    2020/2/4 & 3887  & 65    & 24324 & 490   & 3971  & 23260 & 3156  & 65    & 16678 & 479 \\
    2020/2/5 & 3694  & 73    & 28018 & 563   & 5328  & 24702 & 2987  & 70    & 19665 & 549 \\
    2020/2/6 & 3143  & 73    & 31161 & 636   & 4833  & 26359 & 2447  & 69    & 22112 & 618 \\
    2020/2/7 & 3399  & 86    & 34546 & 722   & 4214  & 27657 & 2841  & 81    & 24953 & 699 \\
    2020/2/8 & 2656  & 89    & 37198 & 811   & 3916  & 28942 & 2147  & 81    & 27100 & 780 \\
    2020/2/9 & 3062  & 97    & 40171 & 908   & 4008  & 23589 & 2618  & 91    & 29631 & 871 \\
    2020/2/10 & 2478  & 108   & 42638 & 1016  & 3536  & 21675 & 2097  & 103   & 31728 & 974 \\
    2020/2/11 & 2015  & 97    & 44653 & 1113  & 3342  & 16067 & 1638  & 94    & 33366 & 1068 \\
\hline
    \end{tabular}%
\end{table}%

\autoref{Fig-I1} shows the number of total and daily suspected and confirmed cases in mainland China, Hubei, and China other than Hubei. The accumulated number of suspected cases reaches the maximum on the 19th day (Feb.\,8), and then drops rapidly.

\begin{figure}[h!]
\centering
\subfigure[]   {\includegraphics[width=0.48\textwidth]{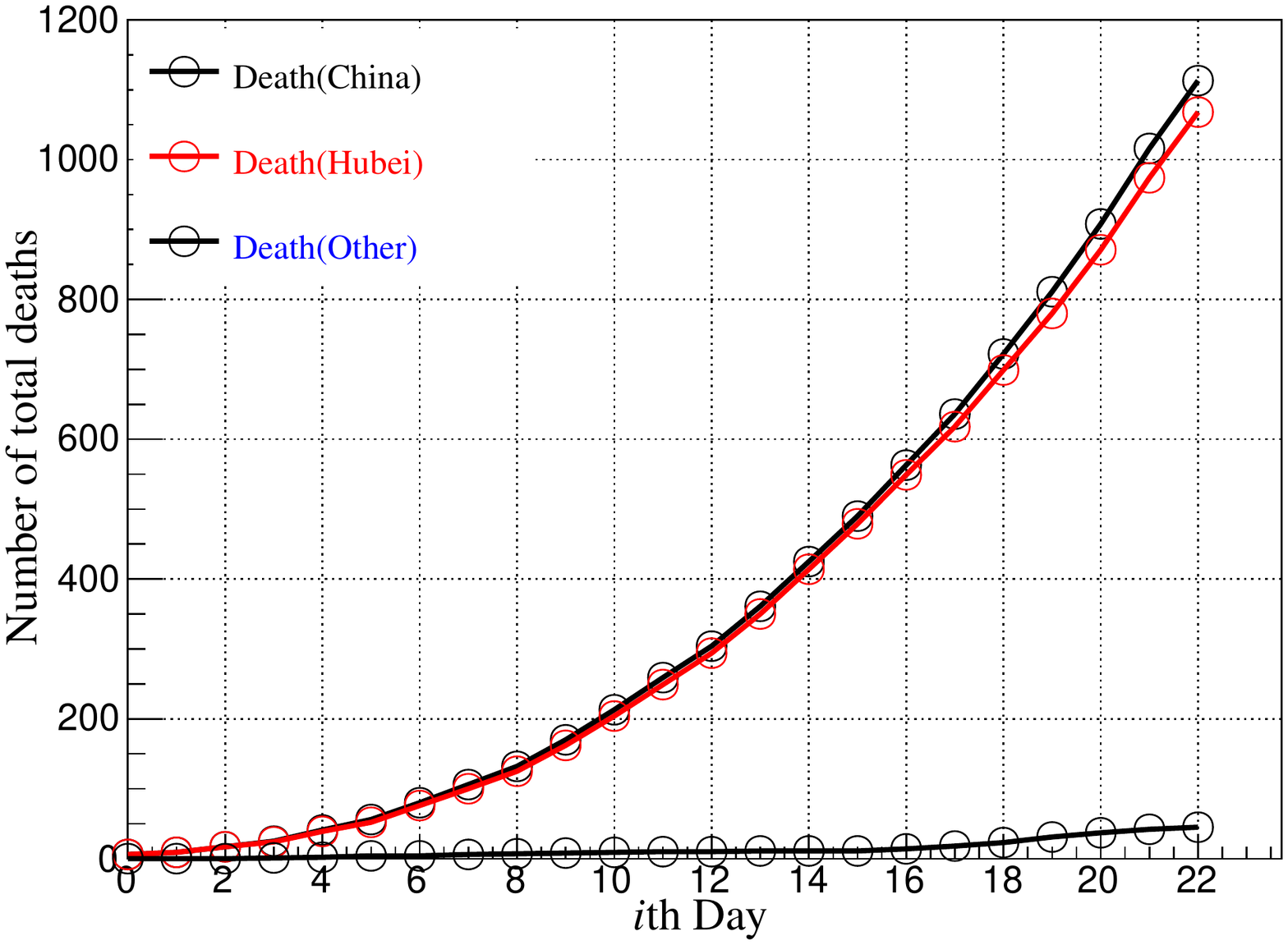} \label{Fig-I2-1}}
\subfigure[]   {\includegraphics[width=0.48\textwidth]{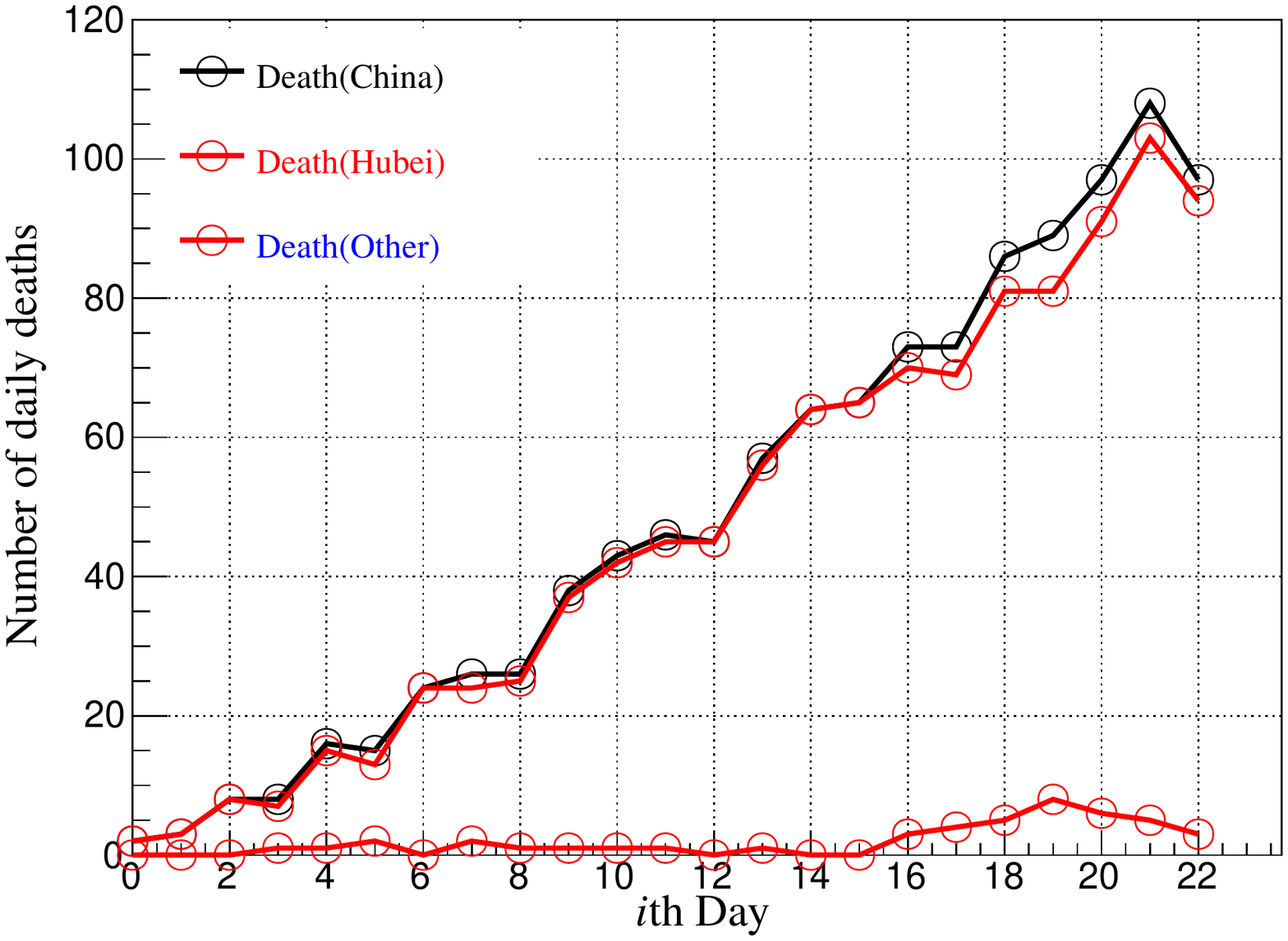} \label{Fig-I2-2}}
\caption{The number of (a) total and (b) daily deaths in mainland China, Hubei, and China other than Hubei.} \label{Fig-I2}
\end{figure}

\autoref{Fig-I2} displays the number of total and daily deaths caused by this epidemic in mainland China, Hubei, and China other than Hubei. Notice the data of the deaths caused by the epidemic in China (black) almost coincides with that in Hubei (red). Almost all the cases of deaths (1068/1113, 96\%,  until Feb.\,11, 2020) locates in Hubei province. This also motivates us to process the data in  Hubei separately with that in China other than Hubei.

\autoref{Fig-I3-2} shows the total and daily deaths over time in China other than Hubei, and the number of total deaths is 45 until Feb.\,11, 2020. The varies of death rate over time is displayed in \autoref{Fig-I3-1}, and the death rate in China other than Hubei is multiplied by a factor of 10. 

\section{Methods}
The key point to describe and predict the trend of the epidemic is to find the proper functions.
We use the symmetrical function $h(t)$ to describe the data of daily infections and deaths in Hubei, namely,
\begin{gather}
h(t)= A\left[\frac{1}{1+e^{-k x}} - \frac{1}{1+e^{-k(x-1)}} \right],
\end{gather}
where $x=(t+0.5-t_\up{T})$ with $t$ denoting the day, which starts from 0 (Jan. 20, 2020) in the data, and $t_\up{T}$ representing the turning point; $A$ and $k$ are the fitting parameters and are determined by the data together with $t_\up{T}$. And then the accumulated data of infections or deaths are determined by the integration over $h(t)$, namely, 
\begin{gather}
H(t) = \int_{-\infty}^{t} h(\tau) \up{d} \tau,
\end{gather}

For the epidemic in the other areas of China, the data of infections shows an asymmetric character, and then will be described by
\begin{gather}
s(t)= B \exp[-(x+k_1 e^{-x/k_1})],
\end{gather}
where $x=t-t_\up{T}$; the parameters $B$, $k_{1}$, and $k_2$ together with $t_\up{T}$, are then determined by fitting to the data. The is function behaves similar with the Gauss function, but has a long tail in the right hand side of the maximum-value location. The corresponding accumulated data can be then obtained similarly with that in Hubei. Though the two models listed here are quite simple, the obtained results in next section show that they would give quite well descriptions to the data of the epidemic.

\section{Results}
\subsection{Predictions of the epidemic in Hubei}

\begin{figure}[h!]
\centering
\subfigure[]   {\includegraphics[width=0.48\textwidth]{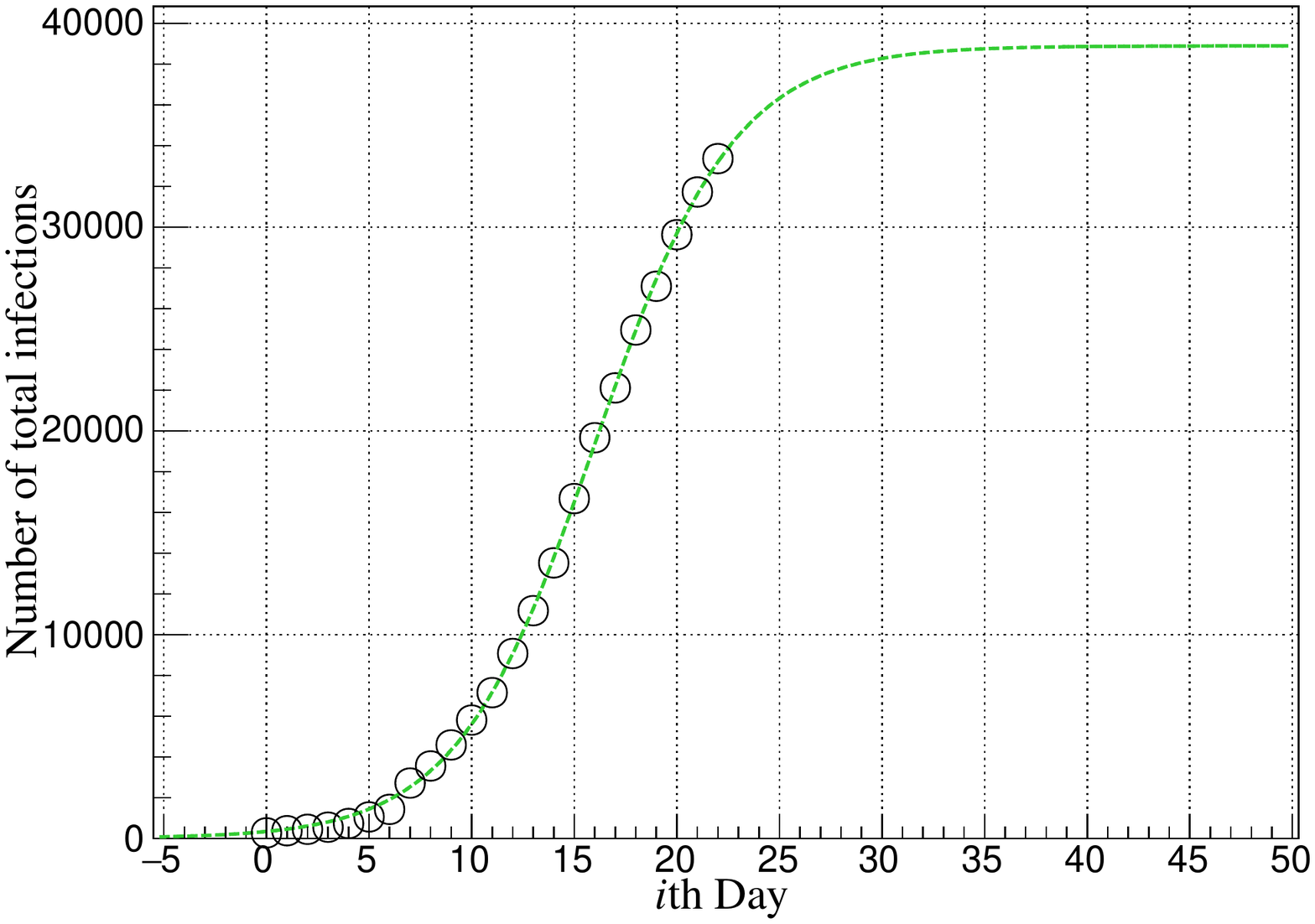} \label{Fig-Hub-In-1}}
\subfigure[]   {\includegraphics[width=0.48\textwidth]{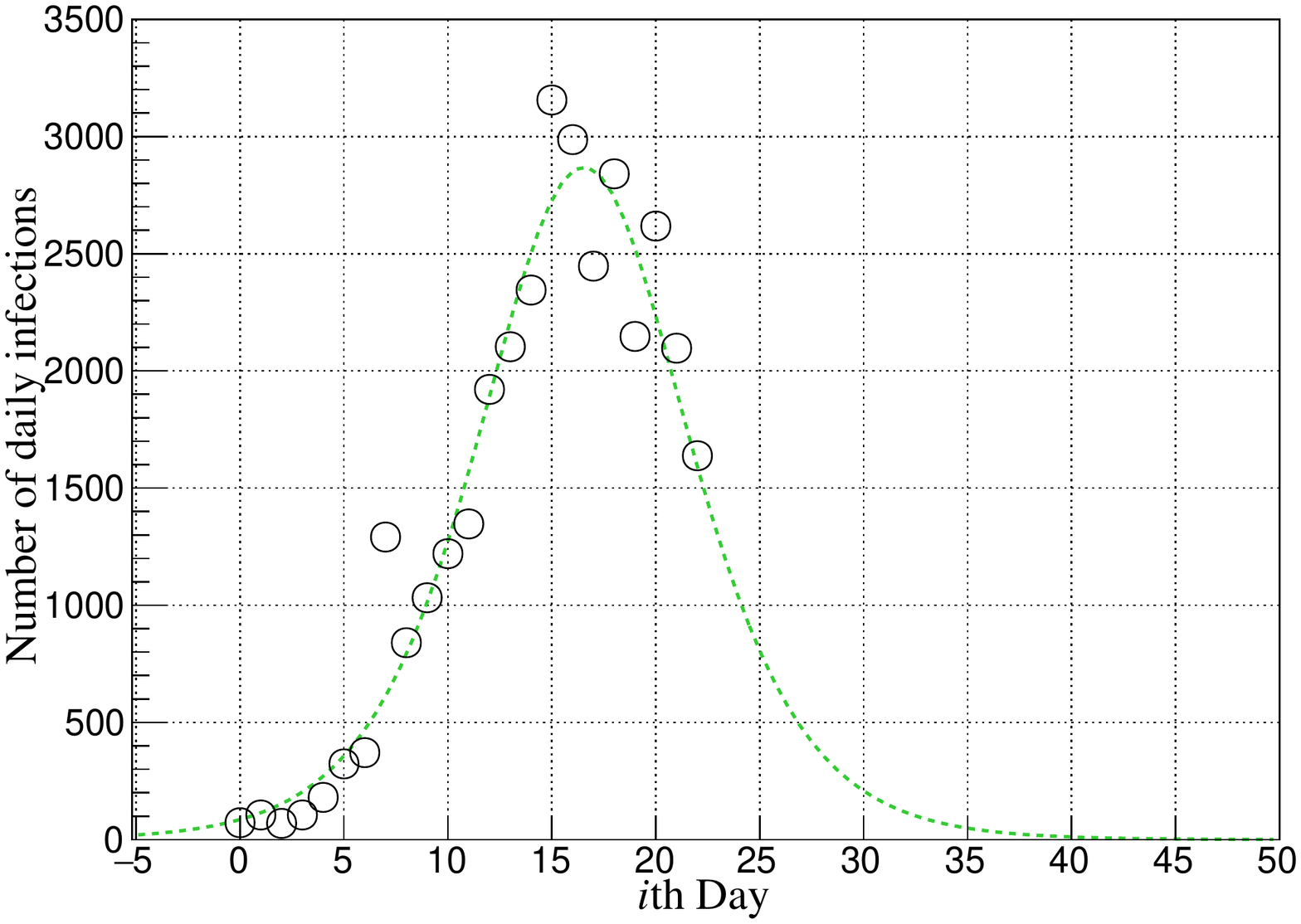} \label{Fig-Hub-In-2}}
\caption{Data and fit of the infections in Hubei of China, with (a) the total infection number, and (b) the daily infection number; the black circle denotes the data, and the green dotted line is the predicted trend; the turning point for infection number in Hubei is calculated to be 17th day, namely, Feb.\,6, 2020.} \label{Fig-Hub-In}
\end{figure}

With considered data, namely, data from Jan.\,20 to Feb.\,11, the average error is bout 166 for this model to describe the daily infections, and 190 for the accumulated total infections. The average relative error is about $8.6\%$ for the number of daily infections during this 23 days, and $1.6\%$ for the number of the accumulated infections.

Our estimates of the number of total and daily infections are showed in \autoref{Fig-Hub-In}. The extracted turning point of the infection number in Hubei is the 17th day, namely, Feb.\,6, 2020. The epidemic in Hubei is predicted to end after the 50th day, namely, after Mar.\,10, 2020. The daily infection number is predicted to be less than 1000, 100, and 10 on the 25th (Feb.\,14), the 33th (Feb.\,22), and the 41th (Mar.\,1) day, respectively. We estimated that the epidemic is to end up with a total of 39, 000 infections in Hubei, China.

\begin{figure}[h!]
\centering
\subfigure[]   {\includegraphics[width=0.48\textwidth]{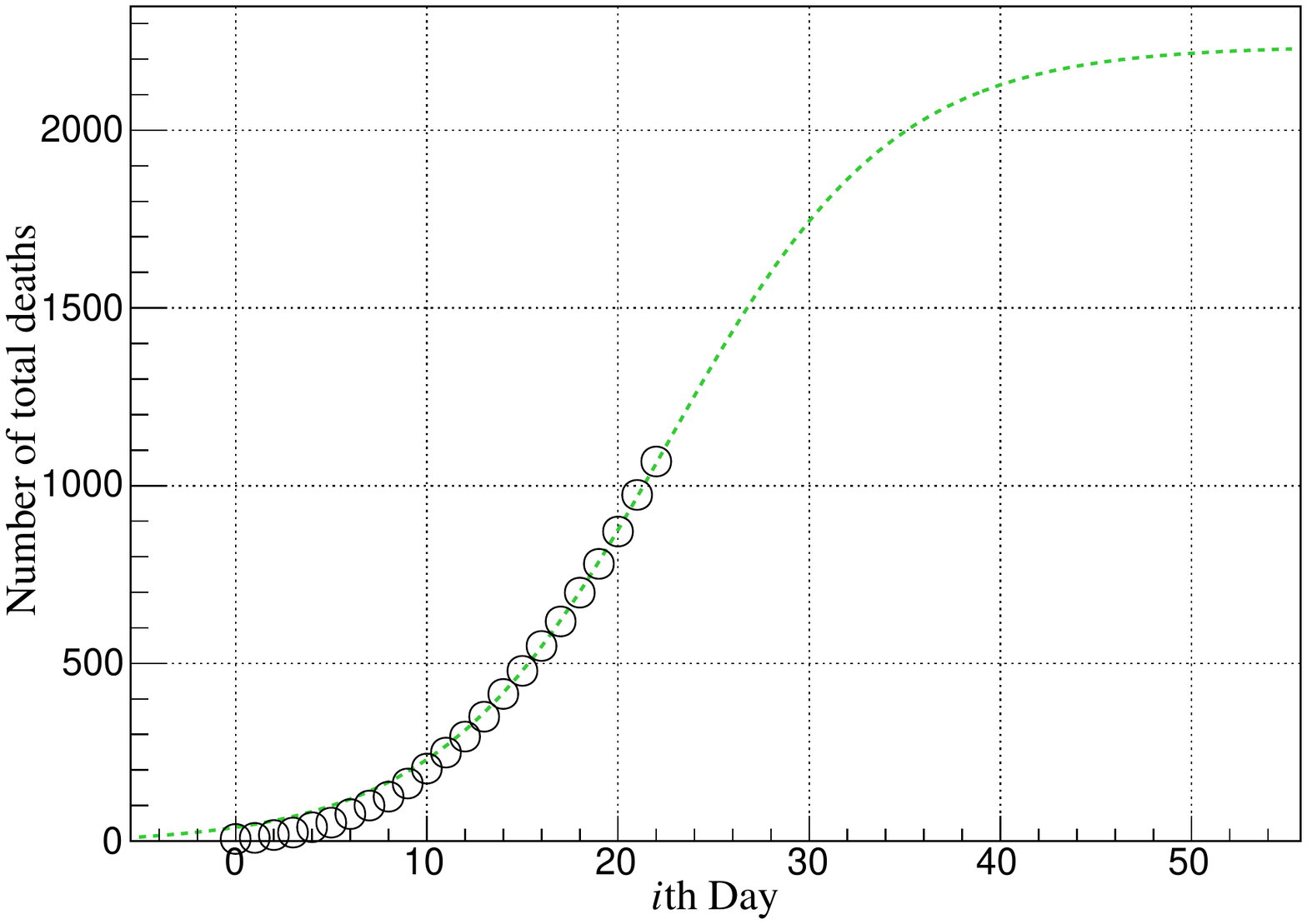} \label{Fig-5-1}}
\subfigure[]   {\includegraphics[width=0.48\textwidth]{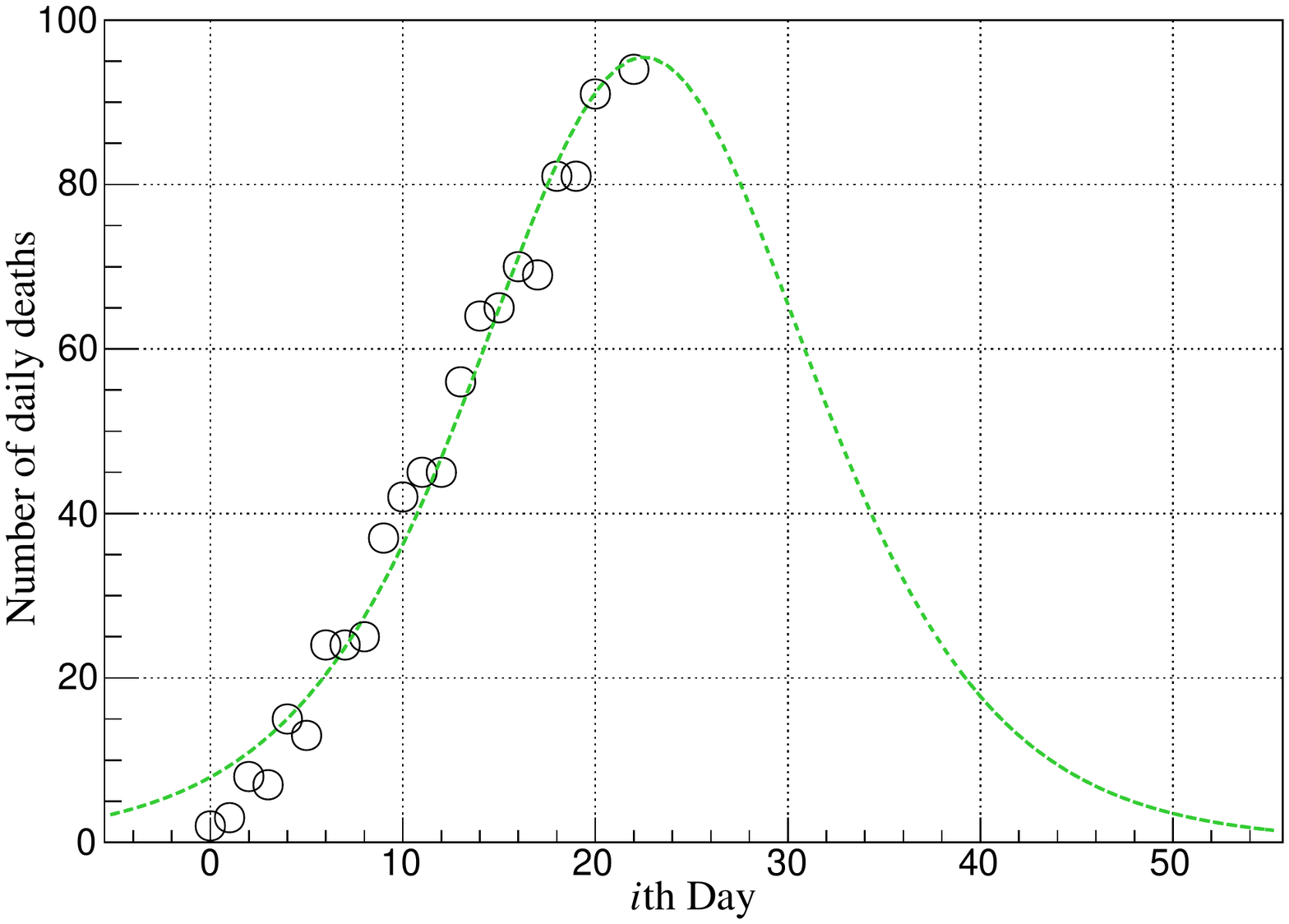} \label{Fig-5-2}}
\caption{Data and fit of the deaths in Hubei of China, with (a) the total death number, and (b) the daily death number; the black circle denotes the data, and the green dotted line is the predicted trend; the turning point for death number in Hubei is calculated to be the 23th day, namely, Feb.\,12, 2020.} \label{Fig-Hub-Death}
\end{figure}

With the considered data, the average errors are bout 4 and 22 for this model to describe the daily and accumulated death numbers, respectively, and the corresponding relative errors are about $8.6\%$ and $6.2\%$, respectively.

 \autoref{Fig-Hub-Death} shows the estimations of the total and daily death number in Hubei, China. The predicted turning point is the 23st day (Feb.\,12, 2020) with the maximum daily death to be less than 100 individuals. The daily death number is estimated to be less than 80, 40, and 10 on the 28th (Feb.\,17), 35th (Feb.\,24), and 44th (Mar.\,4) day, respectively. Notice the distribution of the daily deaths is delayed about $5\sims6$ days compared with the that of the daily infections.

\subsection{Predictions of the epidemic in China other than Hubei}

With the data in the considered period, the average errors are bout 41 and 58 for this model to describe the daily and total accumulated infections, and the corresponding relative errors are about $8.4\%$ and $1.2\%$, respectively. 

\begin{figure}[h!]
\centering
\subfigure[]   {\includegraphics[width=0.48\textwidth]{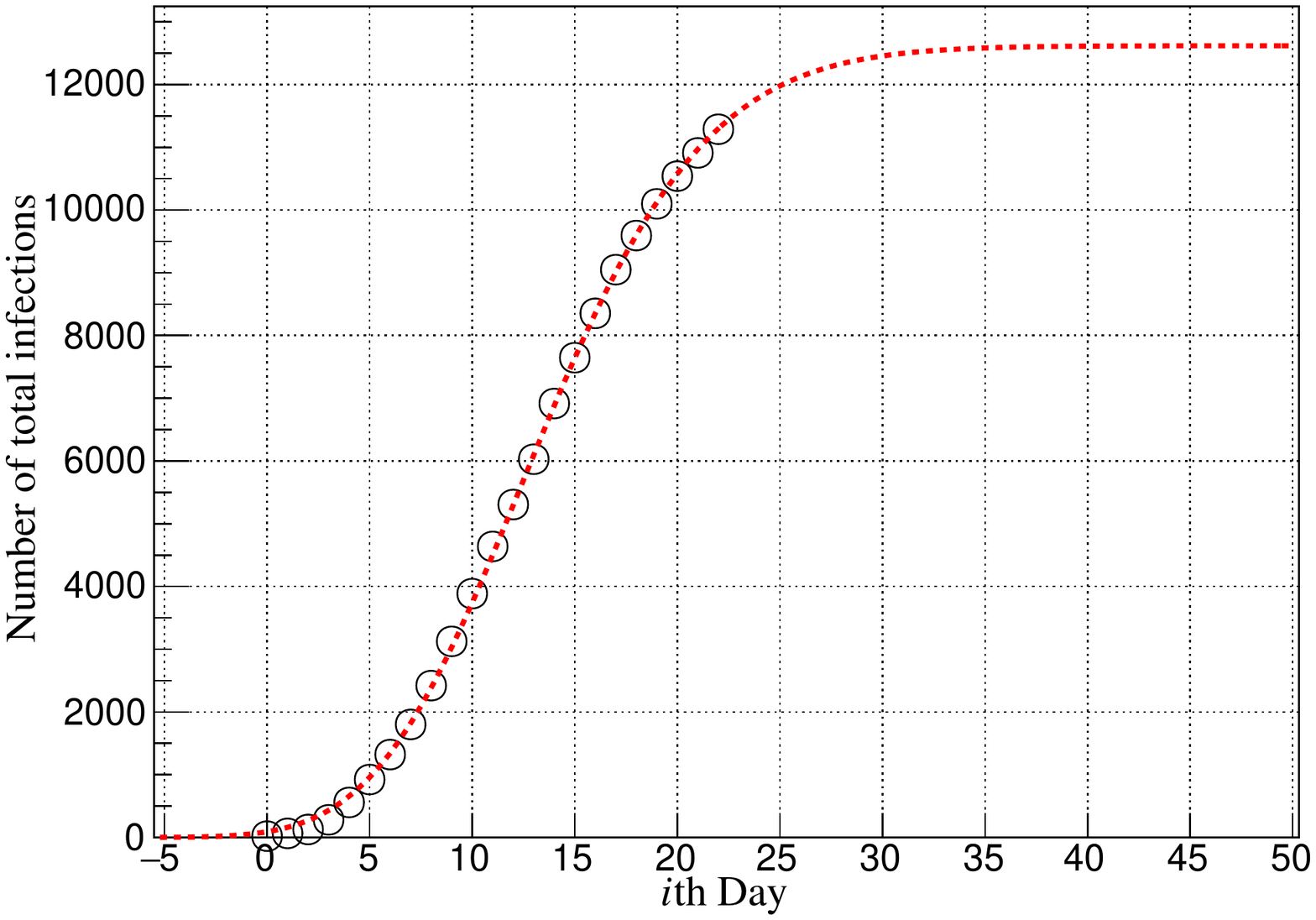} \label{Fig-Nhb-In-1}}
\subfigure[]   {\includegraphics[width=0.48\textwidth]{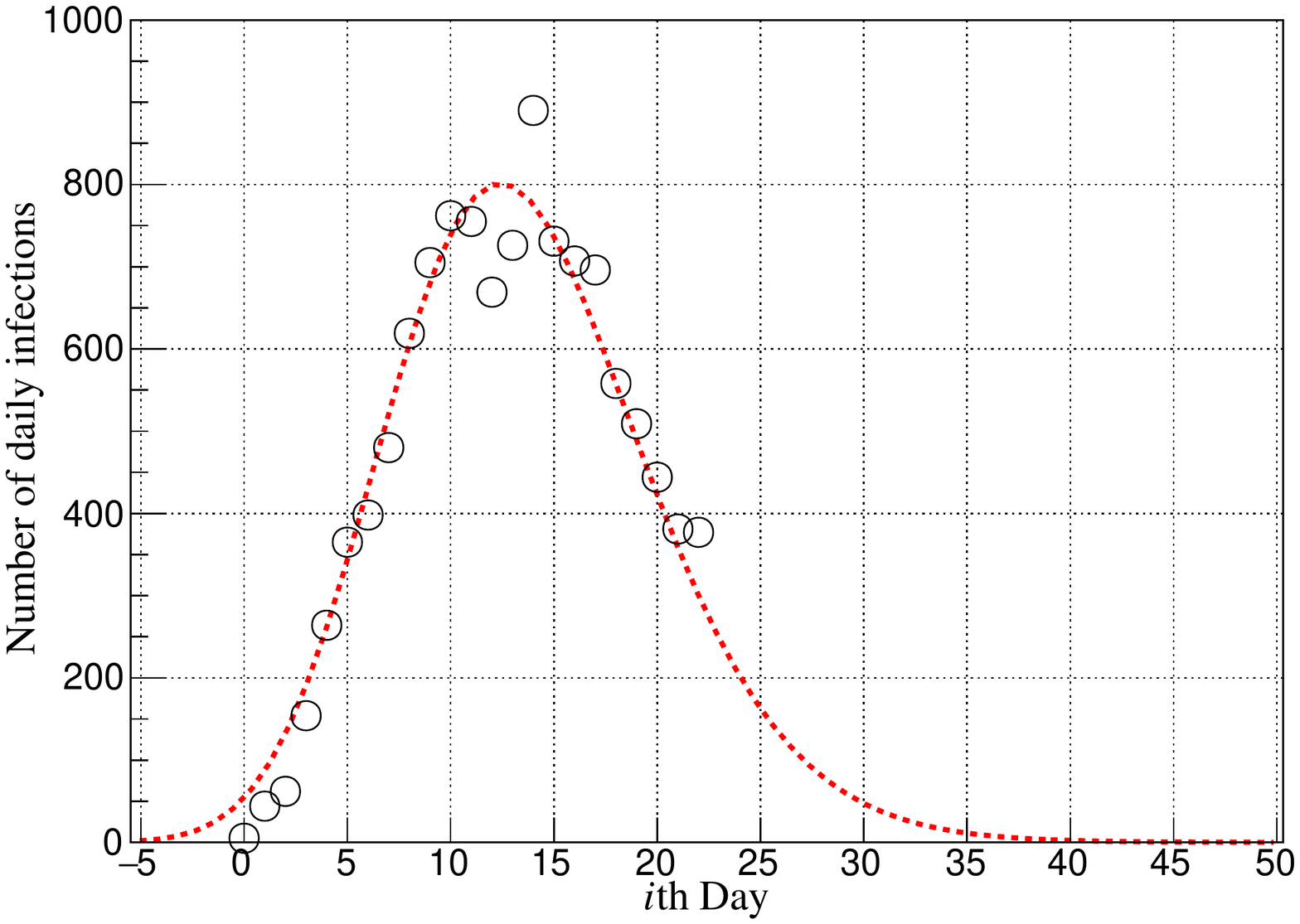} \label{Fig-Nhb-In-2}}
\caption{Data (black circles) and fit (dotted line) of the infections in China other than Hubei, with (a) the total infection number, and (b) the daily infection number; the turning point for infection number is calculated to be the12th day, namely, Feb.\,1, 2020.} \label{Fig-Nhb-In}
\end{figure}

The numbers of the daily and total infections in China other than Hubei are showed in \autoref{Fig-Nhb-In-1} and \autoref{Fig-Nhb-In-2}, respectively. The extracted turning point is the 12th day (Feb.\,1, 2020). The epidemic in China other than Hubei is expected to end on the 45th day, namely, on Mar. 5, 2020. The daily infection number is predicted to be less than 300, 80, and 10 on the 23th (Feb.\,12), the 28th (Feb.\,17), and the 36th (Feb.\,25) day, respectively. The estimated number of accumulated infections is about 12600 in China other than Hubei.

Due to the minority of the statistical data in death number of China other than Hubei, 45 deaths in the last 20 days, we did not parameterize this data, and hence did not give a description or trend prediction.

\section{Summary and discussions}

The data of the daily and total infections and deaths in Hubei and China other than Hubei are studied and parameterized with proper models. The turning points of the daily infections are predicted to be Feb.\,6 and Feb.\,1, 2020, for Hubei and China other than Hubei, respectively. The epidemic caused by the COVID-19 in China is predicted to end up after Mar. 10, and the number of the total infections are predicted to be 51600. The data trends show that the quick and active strategies to reduce human exposure taken in China have already had a good impact on control of the epidemic.

The techniques used here are data-driven and quite different from the work in Ref.\,\cite{WuJT2020}.
The methodology and forecast results here could give some support for the prevention and control of the outbreaks.
The model in this work is parameterized with the latest data until Feb.\,11, 2020, reported by the NHC and HCH.
A major limitation of this work is that we describe the epidemic data in Hubei with the symmetrical even function. Usually, the epidemic would recede with a long tail, and we might underestimate the total infections and deaths in Hubei, and also the end date.
Another limitation is that although we process separately the data of Hubei and China other than Hubei, this is still quite rough for the considerable differences among cities within Hubei province. Also, as a data-driven forecast study, the seasonality of COVID-19 transmission, the effects of the quarantine, the limitations on population transportation, and other specific reasons that may affect the spread of the epidemic, are not explicitly included in the study.  At last, we expect this outbreak would end as soon as possible.

\acknowledgments
We thank Jing Li and Hao-Nan Wang for the helpful discusses and suggestions. This work is supported by the Open Research Fund of Key Laboratory of Digital Earth Science (2019LDE005), and by the Fundamental Research Funds for the Central Universities under Grant No.\,310201911QD054.

\paragraph{Conflicts of Interest:} The authors declare no conflict of interest.

\paragraph{Note added.} After this work has been written, the Hubei reported 14840 confirmed infections (including 13332 clinically diagnosed cases) on Feb.\,12, 2020, which is almost 9 times greater than the data of the previous day. This huge fluctuation of the infections is due to the changing of diagnostic criteria in Hubei. And we believe this clinical criteria taken by Hubei would play an active and important role in controlling the outbreak. 

In order to continue this forecasting research, we consider these 13332 clinically diagnosed infections follow the same statistical distribution as these cases confirmed by nucleic acid detection. Then the updated prediction of final accumulated infections would be 1.4 times larger than before, namely,  totally there would be 54600 infections in Hubei. However, this would overestimate the real COVID-19 infections, since part of the clinically diagnosed cases would be excluded by the nucleic acid detection later.

% The bibliography will probably be heavily edited during typesetting.
% We'll parse it and, using the arxiv number or the journal data, will
% query inspire, trying to verify the data (this will probalby spot
% eventual typos) and retrive the document DOI and eventual errata.
% We however suggest to always provide author, title and journal data:
% in short all the informations that clearly identify a document.
%\bibliographystyle{../../articles/BST-QIANG}
%\biboptions{numbers,sort&compress}
\setlength{\bibsep}{0.8ex}  % vertical spacing between references
%\bibliography{../../articles/reference-QIANG}

\end{document}